\documentclass{article}
\usepackage{
spconf,
commath,
amsmath,
amssymb,
geometry,
graphicx,
hyperref,
booktabs,
breqn,
rsfso
}

\usepackage[edges]{forest}
\usetikzlibrary{shadows}
\usepackage[symbol]{footmisc}
\usepackage{tabularx}

\title{TTSDS - Text-to-Speech Distribution Score}
\name{Christoph Minixhofer\thanks{PhD funded by Huawei Technologies (UK) Co., Ltd.\newline and supported by Google TPU Research Cloud (TRC).}, Ondřej Klejch, Peter Bell}
\address{The University of Edinburgh\\Centre for Speech Technology Research}
\begin{document}
\ninept
\maketitle
\begin{abstract}
Many recently published Text-to-Speech (TTS) systems produce audio close to real speech. However, TTS evaluation needs to be revisited to make sense of the results obtained with the new architectures, approaches and datasets. We propose evaluating the quality of synthetic speech as a combination of multiple factors such as prosody, speaker identity, and intelligibility. Our approach assesses how well synthetic speech mirrors real speech by obtaining correlates of each factor and measuring their distance from both real speech datasets and noise datasets. We benchmark 35 TTS systems developed between 2008 and 2024 and show that our score computed as an unweighted average of factors strongly correlates with the human evaluations from each time period.
\end{abstract}

\section{Introduction}
\label{sec:related}

There has been a recent surge in synthetic speech generation quality enabled by systems using language modeling architectures generating discrete units~\cite{defossez2022encodec} which are then used to reconstruct the waveform \cite{lyth2024natural}.
However, many of the most recent systems have been released to the community without accompanying papers and/or evaluation. This is understandable since the field is moving at a quick pace, and evaluation of synthetic speech is hard.
The remaining systems are evaluated mostly with Mean Opinion Score (MOS) and its factors such as naturalness, speaker similarity, and sometimes intelligibility. Some systems also rely on MOS predicted with trained neural networks \cite{chen2024vall,casanova2024xtts}. Unfortunately, MOS is becoming less useful as real and synthetic speech get closer in quality, and cannot be compared across studies and over time \cite{le2022back}.

Many factors, such as prosody, intelligibility, naturalness, and speaker similarity, contribute to perceived overall speech quality.
Some works focus on the prosody factor by comparing pitch or duration of real to synthetic speech \cite{budzianowski2024pheme,li2024styletts,minixhofer2023eval}, while others include metrics on Word Error Rate (WER) to account for intelligibility or more general algorithmic measures such as Mel-Cepstral Distoration (MCD) \cite{chen2024vall,budzianowski2024pheme}.
To the best of our knowledge, there is only one published effort which evaluated state-of-the-art TTS systems based on LLM architectures \cite{ttsarena}. It uses crowdsourced A/B preference tests to produce an Elo rating for each system. As these ratings are the first comparative study of the next generation of TTS systems, they are a great resource, however this type of evaluation comes with a set of challenges. Since the evaluation is centralized, it is up to the organizers to add newly released systems. Furthermore, human rating of the speech could change as time progresses, which has been shown to be the case for MOS scores \cite{le2022back}. 
Finally, adding more complex domains such as long-form speech synthesis, or testing the contribution of individual factors, requires running the evaluation from scratch.

Some solutions for objective evaluation of generative systems have emerged in other domains: In image generation, Fréchet Inception Distance (FID) \cite{heusel2017fid} has become the de-facto standard, but while there have been attempts to apply these methods to TTS \cite{gritsenko2020fdsd}, they have not taken hold, perhaps due to the large number of required samples~\cite{chen2023pfid}.
In NLP, the advances of capabilities of Large Language Models (LLMs) have lead to a number of published benchmarks, most ranking the model on a variety of tasks, such as the GLUE \cite{wang2018glue}, SuperGLUE \cite{wang2019superglue} and CoQA \cite{reddy2019coqa} benchmark. For speech processing, the SUPERB benchmark also uses this task approach \cite{yang2021superb}.

Unlike math problems, or automatic speech recognition (ASR), speech synthesis does not have one correct solution, but many possible ones, which makes evaluation difficult. However, we can use the concept of factors
for evaluation. As we can test a LLM on mathematical ability and reading comprehension as parts of an overall measure of reasoning ability, we can test a TTS system on factors such as prosody or intelligibility, and we can combine them to reflect the overall performance of the system. This approach also gives us more information on preferring one system over another. For example, for character voices in a video game, a system with higher scores in prosody might be preferred, while for a language-learning app intelligibility would be the most important factor.

In this paper, we devise a benchmark with the \textit{intelligibility}, \textit{prosody}, \textit{speakers}, \textit{environment} and \textit{general} factors. We include \textit{intelligibility} and \textit{prosody} as they are important measures of synthetic speech quality \cite{campbell2007eval}. We include the \textit{speaker} category to evaluate how closely TTS systems can model realistic speakers \cite{fan2015multi}. We also include \textit{environment} as a factor due to the prevalence of artifacts present in speech synthesis \cite{wagner2019eval}, and the difficulty of some TTS system to generate speech with realistic recording conditions \cite{minixhofer2023eval}.
The \textit{general} factor is similar to previous measures of speech distributions as represented by latent representations such as Fréchet DeepSpeech Distance \cite{gritsenko2020fdsd}.
Finally, we compute the overall TTSDS score as an average of all factors.

We compute the score for each factor by comparing the distributions of both high-dimensional features (e.g., embeddings) and scalar features (e.g., pitch) extracted from the speech. This comparison allows us to measure the deviation of synthetic speech from real speech without assuming predefined distributions, thereby avoiding common pitfalls associated with objective speech evaluation measures. For instance, intelligibility is often quantified using Word Error Rate (WER) \cite{cooper2024ttseval}, where lower values are typically preferred. However, if the target domain naturally exhibits high WER (e.g., children's speech), a TTS system should also reflect this characteristic, producing higher WER accordingly. Therefore, we compare the utterance-level distribution of our features (such as WER) rather than their mean.

We evaluate our benchmark by comparing to MOS scores and A/B test results obtained for 35 TTS systems from 2008, 2022 and 2024. An average of our factor scores show correlation coefficients ranging from 0.60 to 0.83 for each time period, while the performance of state-of-the-art MOS prediction are less consistent, ranging from 0.05 to 0.85. Additionally, we observe a shift in priorities of human evaluators over time, with \textit{environment} being more important for earlier systems, and \textit{prosody} being more important for later ones. We make our benchmark suite and leaderboard openly available.\footnote{\url{https://ttsdsbenchmark.com/leaderboard}}

\section{Methodology}
\label{sec:method}
The first step of developing any measure is to define the concept being measured. Since there is no objective way of directly measuring the naturalness or quality of synthetic speech, we define our measure as "the distance between the distribution of real and synthetic speech". The next step is to define relevant factors \cite{viswanathan2005measuring}. For speech we define the following five major factors.

A \textit{General} factor which measures the distribution of the speech without any assumptions. We use self-supervised learning (SSL) representations of the speech.

An \textit{Environment} factor which measures noise or distortion present in the speech. We use correlates for signal-to-noise ratio (SNR) and reverberation.

An \textit{Intelligibility} factor which measures how easy the content of the speech is to recognize. We use WER obtained using the transcripts provided to the TTS systems and ASR systems.

A \textit{Prosody} factor which measures how realistic the prosody of the speech is. We use a pitch extractor, a SSL representation of the prosody and a proxy for duration and speaking rate.

A \textit{Speaker} factor which measures how close the speakers are to real ones. We use representations obtained from speaker verification systems.

\begin{table}[t]
  \centering
    \begin{tabularx}{0.95\columnwidth}{lX}
        \textbf{Factor} & \textbf{Feature} \\
        \hline
        \textbf{Environment} &    \textbf{Noise/Artifacts} \\ 
        & VoiceFixer \cite{liu2021voicefixer} + PESQ \cite{rix2001pesq} \\
        & WADA SNR \cite{kim2008wada} \\
        \hline
        \textbf{Speaker} & \textbf{Speaker Embedding} \\
        & d-vector \cite{wan2018dvec} \\
        & WeSpeaker \cite{wang2023wespeaker} \\
        \hline
        \textbf{Prosody} & \textbf{Segmental Length} \\
        & Hubert \cite{hsu2021hubert} token length \\
        & \textbf{Pitch} \\
        & WORLD \cite{morise2016world} \\
        & \textbf{SSL Representations} \\
        & MPM \cite{mpm} \\
        \hline
        \textbf{Intelligibility} & \textbf{ASR WER} \\
        & wav2vec 2.0 \cite{baevski2020wav2vec2} \\
        & Whisper \cite{radford2023whisper} \\
        \hline
        \textbf{General} & \textbf{SSL Representations} \\
        & Hubert \cite{hsu2021hubert} \\
        & wav2vec 2.0 \cite{baevski2020wav2vec2} \\
        \hline  
    \end{tabularx}
    \caption{Features used in the benchmark and their respective factors. The overall TTSDS score is computed as an average of individual factor scores.}
    \label{tab:features}
\end{table}

Since none of these factors can be measured directly, we rely on several features which correlate to the factors. The extensive body of work in speech processing and representation learning provides these features, including representations from self-supervised models, statistical features, and algorithmic features. For each feature derived from synthetic data, we define its \textit{score} as how close it is to the same feature derived from real speech.

\subsection{Features extracted from Speech}
\label{sec:features}


Here we identify the specific features (i.e., data points derived from speech) that represent each factor. Table~\ref{tab:features} summarizes the models and algorithms that we use to extract these features. For each of the aforementioned factors, we use two to three features to achieve a robust benchmark despite the low number of 80-100 samples per system.
For measuring the \textit{General} factor score, we use frame-level self-supervised representations extracted from the middle layers of the Hubert base \cite{hsu2021hubert} and wav2vec 2.0 base ~\cite{baevski2020wav2vec2} models.
For the \textit{Environment} distribution, we use two one-dimensional correlates of noise present in the signal -- we use VoiceFixer \cite{liu2021voicefixer} to remove noise from the speech, and then measure PESQ between the enhanced sample and the original one; we also use WadaSNR \cite{kim2008wada} to estimate the SNR of each sample.
For \textit{Intelligibility}, we calculate WER using the reference transcripts and automatic transcripts generated using wav2vec 2.0~\cite{baevski2020wav2vec2} fine tuned on 960 hours of LibriSpeech \cite{panayotov2015librispeech}
and Whisper (small)~\cite{radford2023whisper}. \textit{Prosody} is quantified using frame-level representations from a self-supervised prosody model \cite{mpm} and frame-level pitch features \cite{morise2016world}. 
Additionally, we get a proxy for the segmental durations by using Hubert tokens (with 100 clusters) and extracting their lengths (i.e. how many of the same token occur in a row). Finally, for the \textit{Speaker} factor, we use d-vectors \cite{wan2018dvec} and the more recent WeSpeaker \cite{wang2023wespeaker} representations.

\subsection{Speech Distributions}
\label{sec:sdists}
The distribution of a feature can be computed on any audio dataset, whether it consists of synthetic speech, real speech, or noise. 
We let \( D \) represent an audio dataset, and \( X \) be the feature derived from the dataset. We can sample observed values $x_i$ from the empirical distribution \( \hat{P}(X|D) \) as:
\[
 x_i \sim \hat{P}(X|D)
\]
where \( \mathbf{x}_i \) can be a scalar for one-dimensional features or a vector for multi-dimensional features.


\subsection{Computing Distances Between Distributions}
\label{sec:wasserstein}

To compare the distributions of features derived from different datasets, we utilize the Wasserstein distance, specifically the 2-Wasserstein distance $W_2$, also known as the Earth Mover's Distance. $W_2$ measures the amount of "work" needed to transfer one probability distribution to another as an optimal transport problem \cite{vaserstein1969wasserstein}. This distance measure is widely used in computer vision as the Fréchet Inception Distance (FID)~\cite{heusel2017fid} and in audio processing as the Fréchet Audio Distance~\cite{kilgour2019fad}. 
Here, we formulate how to compute \( W_2 \) given the empirical distributions of a feature \( X \) computed on datasets \( D_1 \) and \( D_2 \) for both the one-dimensional and multi-dimensional case. We denote the corresponding empirical probability distributions \( \hat{P}(X|D_*) \) as \(\hat{P}_* \).

\subsubsection*{One-Dimensional Case}
In the one-dimensional case, \( W_2 \) can be computed as a function of the ordered samples \cite{kolouri2018sliced}: 

\begin{equation}\label{eq:one}
W_2( \hat{P}_1, \hat{P}_2 ) =
\sqrt{
\frac{1}{n}
\sum_{i=1}^{n}
\left( x_i - y_i \right)^2
}
\end{equation}
where \( \{ x_1, \ldots, x_n \} \) denote sorted samples of \( \hat{P}(X|D_1) \), and \( \{ y_1, \ldots, y_n \} \) denote sorted samples of \( \hat{P}(X|D_2) \) .

\subsubsection*{Multi-Dimensional Case}
We can compute \( W_2 \) distance for two normally distributed multi-dimensional \( \hat{P_1} \) and \( \hat{P_2} \) using their mean vectors \( \mu_1 \) and \( \mu_2 \) and their covariance matrices \( \Sigma_1 \) and \( \Sigma_2 \) \cite{heusel2017fid}:


\begin{equation}\label{eq:multi}
W_2( \hat{P}_1, \hat{P}_2 ) =
\sqrt{
\norm{\mu_1-\mu_2}^2+D_{B}(\Sigma_1,\Sigma_2)
}
\end{equation}
where \( D_B \) is the unnormalized Bures metric defined as
\[
D_B(\Sigma_1,\Sigma_2)=
\text{trace} \left (
    \Sigma_1 + \Sigma_1
    - 2 ( \Sigma_2^{1/2} \Sigma_1 \Sigma_2^{1/2} )^{1/2} 
\right )
\]
In line with previous work~\cite{heusel2017fid} this approach assumes latent representations of neural networks are normally distributed.

\subsection{Overall Score}
\label{sec:overall}

\begin{figure}[tb]
    \centering
    \centerline{\includegraphics[width=7cm]{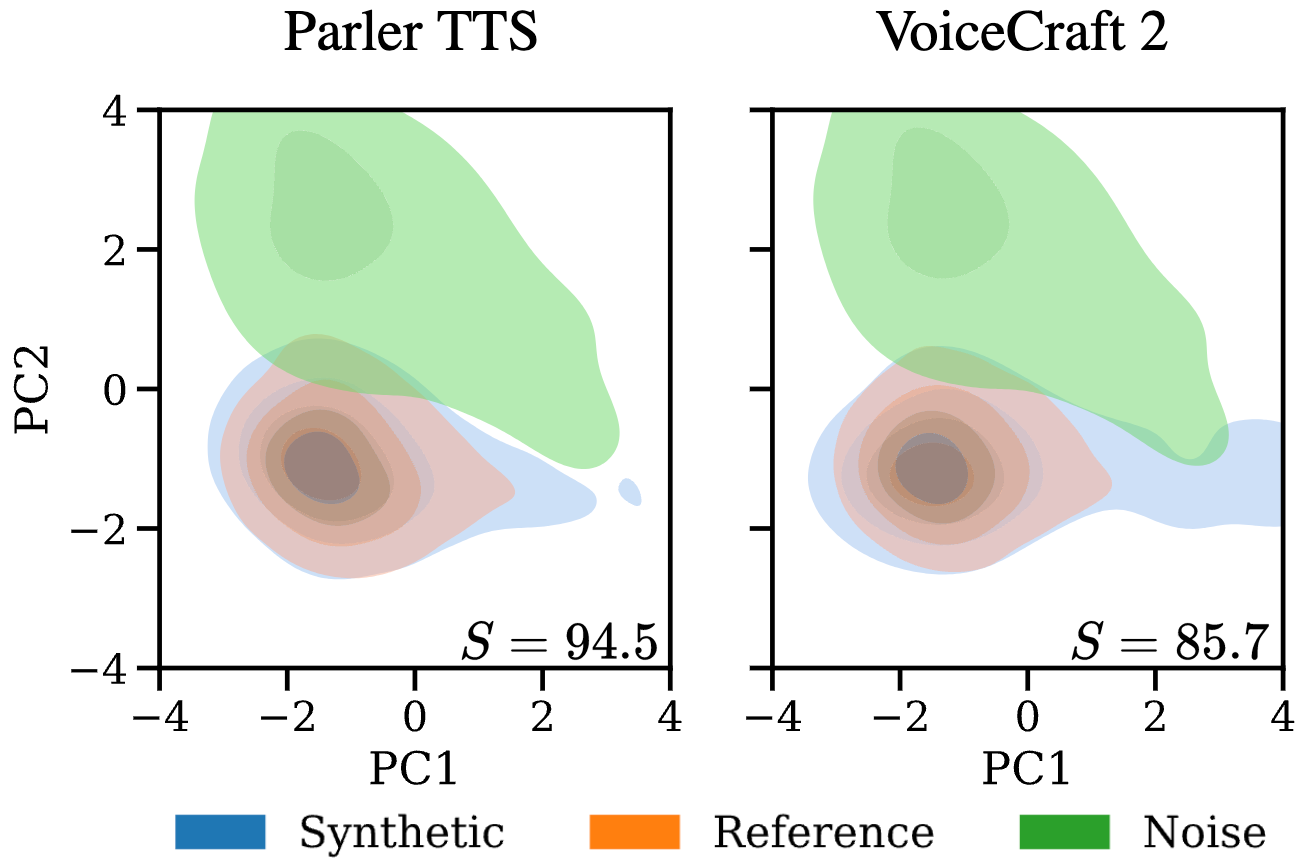}}
	\caption{Distribution of the best (left) and worst (right) TTS Arena system with respect to Hubert representations. \(S\) denotes the score.}
	\label{fig:dist}
\end{figure}

To evaluate how close a synthetic speech dataset \( D_{\text{syn}} \) is to real speech given a particular feature \( X \), we compute its Wasserstein distance \(W_2\) for real reference datasets \( \mathfrak{D}_{\text{real}} \) and distractor (noise) datasets \( \mathfrak{D}_{\text{noise}} \). We find the smallest \( W_2 \) among the real and noise datasets respectively as
\begin{equation*}
  \begin{split}
    W_{\text{real}} &= \min_{D_{\text{real}} \in \mathfrak{D}_{\text{real}}} W_2(\hat{P}_{\text{syn}}, \hat{P}_{\text{real}}) \\
    W_{\text{noise}} &= \min_{D_{\text{noise}} \in \mathfrak{D}_{\text{noise}}} W_2(\hat{P}_{\text{syn}}, \hat{P}_{\text{noise}})  
  \end{split}
\end{equation*}
We define the overall score (ranging from 0 to 100) as
\[
S=100\times \frac{
W_{\text{noise}}
}
{
W_{\text{real}}+W_{\text{noise}}
}
\]
Any score can be intuitively interpreted -- if \( S \) is greater than 50, then \( D_{\text{syn}} \) is more similar to the closest real speech dataset than to the closest noise dataset for a particular feature.
An example of this can be seen in Figure~\ref{fig:dist}, in which we show the difference (for a SSL representation feature) between the best-performing and the worst-performing system in the TTS Arena dataset. The higher score of system (a) corresponds with a larger overlap with the reference dataset and smaller overlap with the noise dataset than system (b).

\section{Experiment Setup}
\label{sec:exp}
\begin{figure*}[t]
	\begin{minipage}[b]{0.25\linewidth}
		\centering
		\centerline{\includegraphics[width=5.4cm]{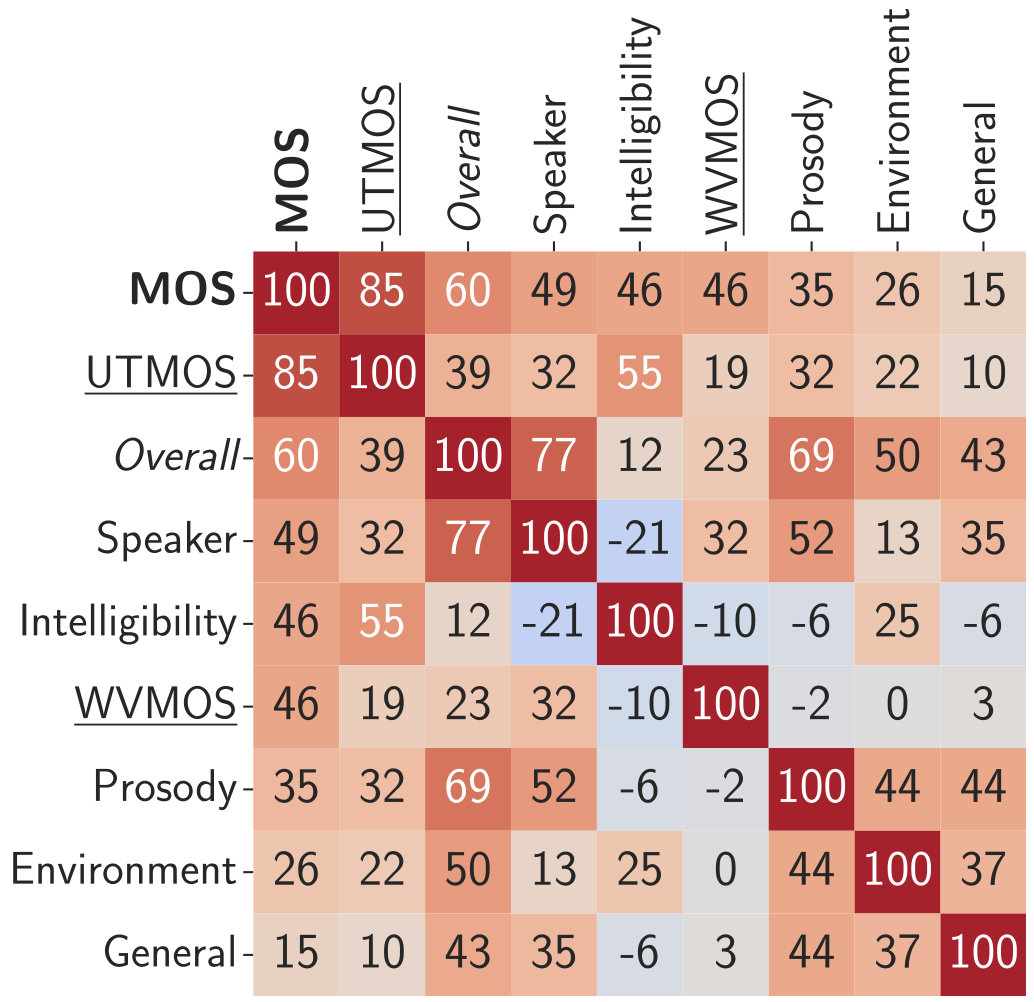}}
		\centerline{(a) Blizzard'08}\medskip
	\end{minipage}
	\hfill
	\begin{minipage}[b]{0.25\linewidth}
		\centering
		\centerline{\includegraphics[width=5.4cm]{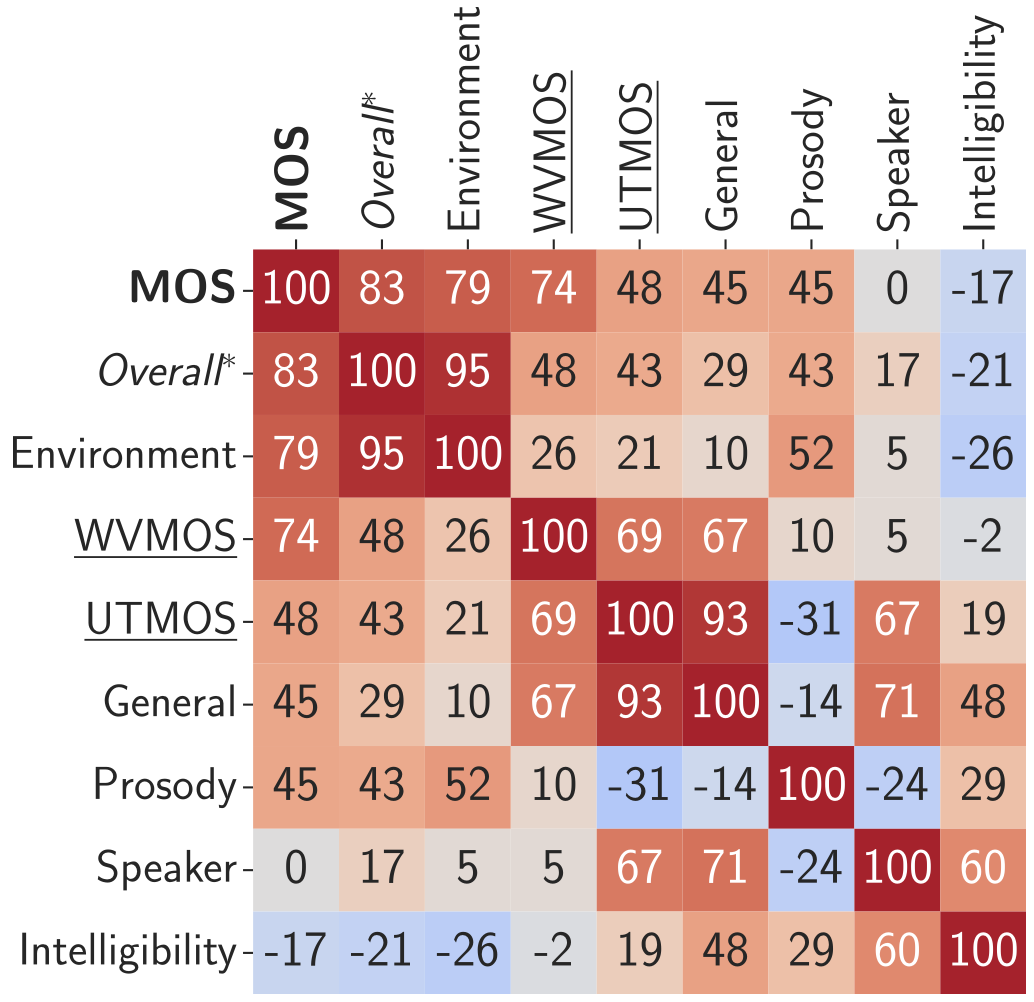}}
		\centerline{(b) BTTF}\medskip
	\end{minipage}
    \hfill
	\begin{minipage}[b]{0.25\linewidth}
		\centering
		\centerline{\includegraphics[width=5.4cm]{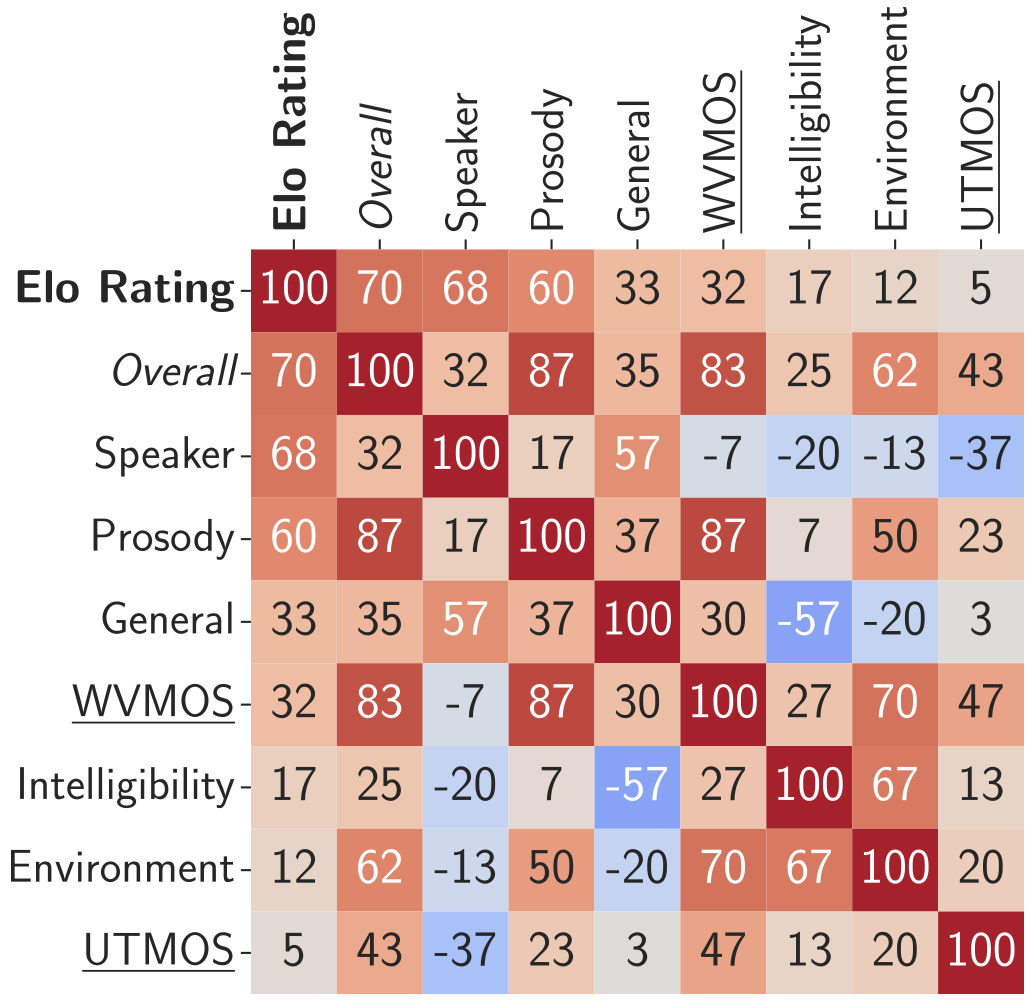}}
		\centerline{(c) TTS Arena}\medskip
	\end{minipage}
    \vspace{-1em}
	\caption{Spearman correlation between the \textbf{subjective measure}, \underline{benchmark systems} and \textit{our benchmark}.}
	\label{fig:heatmap}
\end{figure*}

To validate our benchmark, we compare our factors and TTSDS scores to subjective measures using three different datasets, ranging from legacy to state-of-the-art systems: The Blizzard 2008 challenge \cite{king2008blizzard} compared 22 TTS systems across several tasks using MOS. We choose the "Voice A" audio-book task with 15 hours of data. Later, the "Back to the Future" (BTTF) \cite{le2022back} compared unit selection, hybrid and statistical parametric HMM-based systems from the Blizzard 2013 challenge \cite{king2013blizzard} with deep learning systems inspired by the Blizzard 2021 challenge \cite{ling2021blizzard} based on FastPitch \cite{lancucki2021fastpitch} and Tacotron~\cite{wang2017tacotron}.
The latest systems, which are most commonly based on discrete speech representations generatively modeled by LLM-like systems \cite{chen2024vall}, are represented by the TTS Arena leaderboard \cite{ttsarena}, which is a crowdsourced effort to evaluate these systems. Only systems released in 2023 and 2024 are featured in this evaluation. Since the data is not publicly released, we reproduce datasets for as many of the systems as possible.\footnote{{We had to exclude MetaVoice and GPT-SoVITS due to reference audio length requirements; and MeloTTS due to only female voices being available.}} As conditioning for the TTS systems, we use speaker reference waveforms from the LibriTTS \cite{zen2019libritts} test set, coupled with unrelated transcripts from the same set to make sure the data could not have been encountered during training.

Our reference speech datasets are LibriTTS \cite{zen2019libritts}, LibriTTS-R \cite{koizumi2023librittsr}, LJSpeech~\cite{ljspeech}, VCTK~\cite{vctk}, and the training sets for the Blizzard challenges \cite{le2022back,king2008blizzard}. We sample 100 utterances at random from each dataset (if available, from their test sets).
For distractor/noise datasets, we use the ESC dataset of background noises \cite{piczak2015esc}, as well as the following generated noise -- random uniform, random normal, all zeros and all ones.

We compare our benchmark with two MOS prediction methods. The first method is \textit{WVMOS}~\cite{andreev2022hifi}, which uses wav2vec 2.0 model~\cite{baevski2020wav2vec2} fine-tuned to predict MOS scores. Its system-level correlation coefficients range from 0.68 to 0.96 for different corruptions of speech and their corresponding MOS scores~\cite{andreev2022hifi}.
The second method is \textit{UTMOS}~\cite{saeki2022utmos}, which is an ensemble MOS prediction system that won several categories in the 2022 VoiceMOS challenge \cite{huang2022voicemos}, and has since been used for evaluation of several leading TTS systems \cite{casanova2024xtts,ju2024naturalspeech3}.

For all system\(\times\)feature combinations, we compute the score as described in Section~\ref{sec:overall}.
We average all features for each factor, which gives us the corresponding factor score. Averaging all factor scores in turn gives the TTSDS score.

\section{Results}
\label{sec:results}

Given the scores of all systems and the subjective measures reported for the given datasets (MOS for Blizzard'08 and BTTF; Elo Rating for TTS Arena), we report their Spearman rank correlation coefficients in Figure~\ref{fig:heatmap}.
We observe both our factors and baseline MOS prediction systems vary strongly between different the different corpora, but the TTSDS score correlates consistently well with subjective evaluation results.

The \textbf{baseline MOS prediction methods} achieve mixed results for the Blizzard'08 and BTTF data. UTMOS and WVMOS respectively achieve a high correlation on one of the two datasets while only yielding low correlation on the other. We hypothesize that UTMOS might have included unit-selection systems in its training data, but it have not encountered enough variants of the FastPitch/Tactotron systems present in BTTF. The inverse seems to be the case for WVMOS. For TTS Arena, both systems do not perform well. In summary, these MOS prediction systems sometimes achieve high correlation with ground-truth MOS, but do not seem to generalize.

\subsection{TTSDS Benchmark}
\begin{figure}[t]
    \vspace{-1em}
    \centering\centerline{\includegraphics[width=5.5cm]{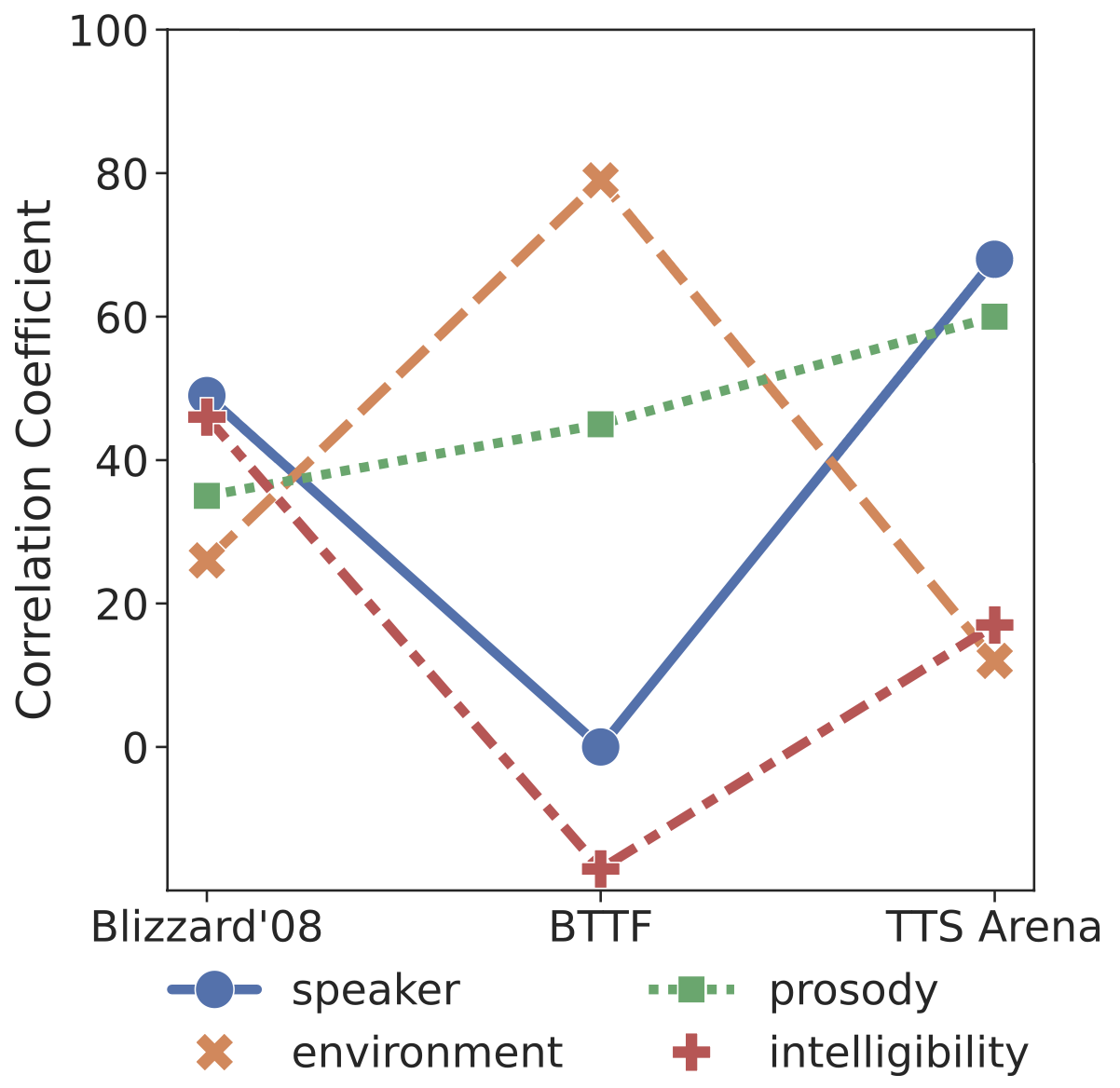}}
    \caption{Development of factor score correlation coefficients over time  from early speech synthesis (Blizzard'08) to the latest systems (TTS Arena).}
    \vspace{-2em}
    \label{fig:overtime}
\end{figure}

We now discuss the individual scores of our benchmark -- the development of individual of these scores' correlations with subjective evaluation can be seen in Figure~\ref{fig:overtime}.

The \textbf{General score} shows some correlation with human evaluation results, but the correlation is generally low. The General score only slightly outperforms MOS prediction for TTS Arena, and shows the lowest correlation of all factors for the Blizzard'08 systems. For unit selection voices, this might be explained by the fact they consist of real speech samples, however, speaker verification representations should suffer from the same problem and they do not seem to be affected as much.

\begin{table*}[t]
\centering
\begin{tabular}{@{}lccccccccc@{}}
\textbf{System} & \textbf{UTMOS} & \textbf{WVMOS} & \textit{Gen} & \textit{Env} & \textit{Int} & \textit{Pro} & \textit{Spk} & \textbf{TTSDS} & \textbf{Elo Rating} \\ \midrule
StyleTTS 2 \cite{li2024styletts}      & \textbf{4.36}                      & 4.48                                & 93.7                                 & 84.7                                     & 91.6                                         & 89.8                                 & 71.5                                 & 86.3                                & \textbf{1237}                           \\
XTTSv2 \cite{casanova2024xtts}          & 3.89                               & 4.36                                & 94.3                                 & 79.3                                     & 91.4                                         & 90.5                                 & \textbf{72.6}                        & 85.6                                & 1232                                    \\
OpenVoice \cite{qin2023openvoice}       & 4.10                               & 4.57                                & 91.7                                 & 88.0                                     & 91.6                                         & \textbf{91.8}                        & 68.8                                 & \textbf{86.4}                       & 1158                                    \\
WhisperSpeech \cite{clapa2024whisperspeech}   & 3.78                               & 3.89                                & 90.0                                 & 83.9                                     & \textbf{92.2}                                & 80.7                                 & 72.4                                 & 83.9                                & 1149                                    \\
Parler TTS \cite{lacombe2024parlertts}      & 3.97                               & 4.16                                & \textbf{94.7}                        & 80.8                                     & 87.5                                         & 83.0                                 & 74.1                                 & 84.0                                & 1140                                    \\
Vokan TTS \cite{buttercream2024vokan}      & 3.80                               & 4.22                                & 88.6                                 & 85.1                                     & 91.6                                         & 85.3                                 & 69.1                                 & 83.9                                & 1126                                    \\
OpenVoice v2 \cite{qin2023openvoice}    & 4.29                               & \textbf{4.75}                       & 90.7                                 & \textbf{91.2}                            & 91.6                                         & 88.6                                 & 68.7                                 & 86.2                                & 1120                                    \\
VoiceCraft 2 \cite{peng2024voicecraft}   & 4.21                               & 3.71                                & 87.0                                 & 78.0                                     & 91.6                                         & 84.4                                 & 66.0                                 & 81.4                                & 1114                                    \\
Pheme \cite{budzianowski2024pheme}          & 3.92                               & 4.26                                & 94.0                                 & 81.9                                     & 91.5                                         & 85.1                                 & 66.1                                 & 83.7                                & 1029                                    \\ 
\end{tabular}
\caption{Ranking, factor scores, TTSDS score and MOS predictions for the TTS Arena systems.}
\label{tab:leaderboard}
\end{table*}

The \textbf{Environment score} has a low correlation with subjective measures for both Blizzard'08 and TTS Arena, but it is interestingly the most important factor for BTTF. Due to BTTF consisting of both deep learning systems from 2021 and systems from 2013, this factor might pick up on artifacts which are only present in the latter.
Meanwhile for the Blizzard'08 systems, these artifacts might be similar enough between systems listeners didn't prioritize them in evaluation, while for modern systems in TTS Arena, hardly any noise or artifacts are present.

The \textbf{Intelligibility score} shows a high correlation for Blizzard'08, but it is the only of our scores showing a negative correlation for BTTF. This could be due to the difference between unit selection and neural voices, with the former perhaps having more realistic intelligibility, but worse naturalness as perceived by humans.

The \textbf{Prosody score} is consistent between datasets, which might be in part due to the diversity of the underlying features (i.e. pitch, SSL prosody representations and segmental durations). It also increases over time, with the TTS Arena system scores showing the highest correlation with our prosody score. This confirms the intuition that good prosody has always been a factor in subjective  evaluation. The increase in prosody score might indicate that human evaluators focus more on the prosody of the speech as other factors such as the intelligibility or noise conditions have vastly improved with modern systems.

The \textbf{Speaker score} also shows high correlation for Blizzard'08 and TTS Arena, but fails for BTTF. We believe this is due to older unit-selection systems included in BTTF producing a natural speaker embedding for concatenated parts of real speech. This effect is pronounced because we only achieve a significant TTSDS score correlation when the \textit{Speech} factor is excluded for BTTF.

\begin{figure}[t]
    \centering\centerline{\includegraphics[width=5.5cm]{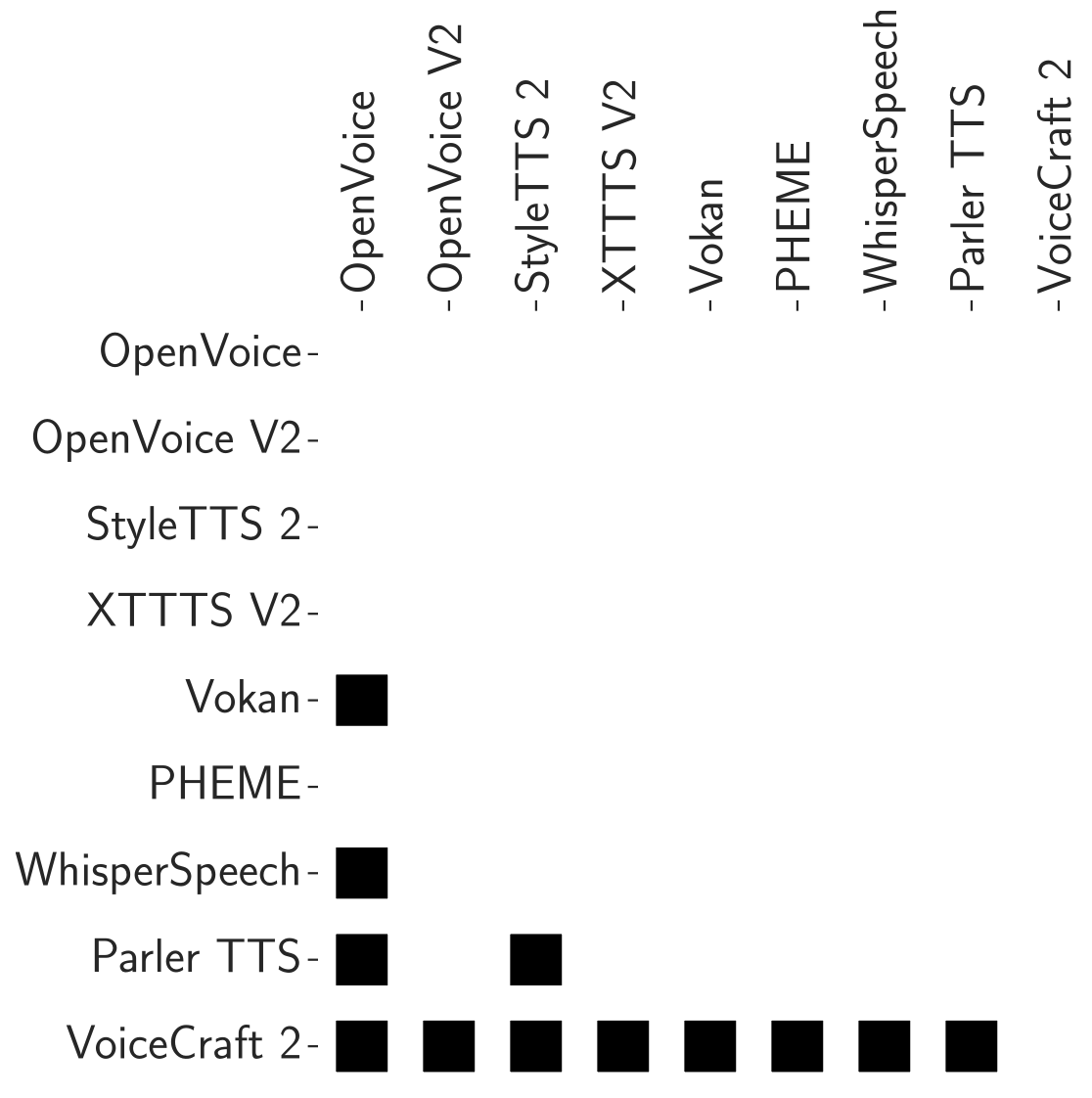}}
    \caption{Results of Wilcoxon signed-rank tests between systems’ extracted features. \(\blacksquare\) indicates a significant difference between a pair of systems.}
    \label{fig:wilcoxon}
    \vspace{-1em}
\end{figure}

The \textbf{TTSDS score} has higher correlation than any single factor for all datasets included in our study, despite the low number of 80-100 samples per dataset. One of the baseline MOS prediction networks still performed better for the early Blizzard'08 systems, but both MOS prediction networks were outperformed by our benchmark for the more modern systems. However, individual factors often show lower correlations with MOS than the baseline systems, showing the need for combining several factors. This might be the reason measures similar to the Fréchet Inception Distance \cite{heusel2017fid} for computer vision have not become popular for speech evaluation -- with the low number of samples typically used for TTS evaluation, and the many factors contributing to what "good" speech synthesis is, and single distance measure might not be enough to show correlation with human evaluation.
Table~\ref{tab:leaderboard} shows our benchmark compared to MOS prediction and the subjective human evaluation rating from TTS Arena. While UTMOS correctly predicts the best system, the other scores by the MOS prediction systems show little to no correlation with Elo ratings; our prosody factor, speaker factor and overall TTSDS scores correlate well with the Elo ratings. However, OpenVoice v2 \cite{qin2023openvoice} is scored highly by TTSDS but achieved low scores in TTS Arena -- this might be due to differences in configuration, as the details for generating the speech used in TTS Arena are not public. 
To evaluate whether our benchmark could be used for system selection, we perform a Wilcoxon signed-rank test (Figure~\ref{fig:wilcoxon}). We observe that while the worst-performing systems can generally be distinguished from the highest-performing ones, there is no statistically significant difference between the better-performing systems. 
Finding significant differences between TTS systems has been difficult, even with previous subjective evaluation methods \cite{le2022back,king2008blizzard}. However, using more speech samples and features for future iterations of TTSDS could mitigate this.



\vspace{-1em}
\section{Conclusion}

In this work, we proposed a benchmark assessing prosody, speaker identity, intelligibility, environment, and general distribution of synthetic speech. Evaluating 35 TTS systems from 2008 to 2024, our benchmark showed strong correlation with human evaluations, showing the robustness and adaptability of our approach to evolving evaluation criteria.
Individual factors alone showed limited correlation, but their combination significantly outperformed traditional MOS prediction systems, especially for modern TTS systems. Our results underscore the importance of intelligibility and prosody, and the need for TTS systems to replicate realistic recording conditions and speaker characteristics.
We revealed limitations in existing MOS prediction systems, emphasizing the need for a nuanced approach to TTS evaluation. High correlation with human evaluations suggests our benchmark provides a reliable and comprehensive framework for assessing synthetic speech quality.



\bibliographystyle{IEEEbib}
\bibliography{refs}

\end{document}